\documentclass[10pt, conference, compsocconf]{IEEEtran}
\ifCLASSINFOpdf
\else
\fi

\usepackage[pdftex]{graphicx}
\graphicspath{{../pdf/}{../jpeg/}{../}}
\DeclareGraphicsExtensions{.pdf,.jpeg,.png}

\usepackage[cmex10]{amsmath}

\usepackage{array}

\begin{document}
%
\title{S.A.T.E.P. : Synchronous-Asynchronous Tele-education Platform}


\author
{
\IEEEauthorblockN{ Lazaros Lazaridis \IEEEauthorrefmark{1},
Maria Papatsimouli   \IEEEauthorrefmark{1} 
and
George F. Fragulis\IEEEauthorrefmark{1}}\\

\IEEEauthorblockA{\IEEEauthorrefmark{1}Laboratory of Web Technologies \& Applied Control Systems\\ Dept. Of Electrical Engineering\\
Technological Educational Inst. of Western Macedonia, 
Kozani, Hellas\\}
}

\maketitle

\begin{abstract}
   S.A.T.E.P. : Synchronous-Asynchronous Tele-education Platform is a  software application for educational purposes, with a lot of parametrizing (configuration) features written entirely from scratch. It aims at the training and examination of  computer skills, a platform that can be adjusted to the needs of each lesson. In the application the trainer and the administrator can define the number of the lectures and upload files for each one of them. Furthermore, he can insert, modify and delete questions which are used for evaluation tests but also for the trainees’ examinations.  The trainee can read/download the files of each lesson and also test his knowledge on what he has studied through a series of questions/tests. A chat module where registered users as well as  system administrator can discuss and solve questions is also developed .

\end{abstract}

\begin{IEEEkeywords}
Web Based Application; MySql; PHP; Open Source Software;Tele-education;  
\end{IEEEkeywords}

%
\IEEEpeerreviewmaketitle

\section{Introduction}
The rapid development of technology, the distance elimination through communication networks and competitiveness have influenced the field of education. Tele-education is a new form of education which is the result of the above factors. In a few words, tele-education aims at education through distance, without the simultaneous presence of the trainers and trainees and the use of new technology such as networks, multimedia etc. It is a form of education which is adjusted to the needs of each trainee and where time and distance don’t matter. On the other side, there is an endless ability of expansion  and there is no restriction in the number of the participants. There are many tele-education platforms either synchronous or asynchronous see for example \cite{Tele1}, \cite{Tele2}. Nevertheless, there are a few platforms that can support the evaluation of trainees automatically.  The S.A.T.E.P. application was designed  as an autonomous platform for synchronous/ asynchronous tele-education  through which the administrator just inserts the various files and questions  and the platform  produces the tests randomly. The application was created under the philisophy of open source software, which gives the potential to anyone having the basic knowledge of programming languages, to adjust the platform to his needs with no cost see also \cite{Skordas_Fragulis_Triant2011} and \cite{Skordas_Fragulis_Triant2014} . We make use  of HTML, JavaScript, MySQL, Apache server, PHP and CSS. The HTML was used for the creation of the web site design\cite {web2008} . JavaScript was used mainly for the control of the elements existing in the forms and generally whenever there is a dynamic content for the local process of the data (client-side) \cite{Java1}. MySQL was used for the design of database.  MySQL is a very powerful tool for the management of database, through which is given the potential to insert, modify and delete tables, data and relations among tables \cite{mysql}. Through MySQL server, is ensured that only authorized users have access to it and that a number of users can work simultaneously on the same database. Apache is an open source web server \cite{apache}. It’s highly adaptive with advanced attributes. Through CSS were created the menus of the web site, for the formatting of the various elements and also for their behavior in certain events. Finally, PHP is a server side script language which is available for all operating systems \cite{php1}. For the creation of the source code of the web site were also used open source programs such as CSSED and tsWebEditor. The S.A.T.E.P. application  is  used in the “Web Programming Language” course of the M.Sc. Applied Informatics course in Kozani .

\section{User Interface}

 In the  S.A.T.E.P. application three types of users are supported. There are the guest users, who don’t have access to the system, the registered users who do have access to the platform and finally administrators. 
\begin{figure}[!h]
	\centering
	\includegraphics[scale=0.3]{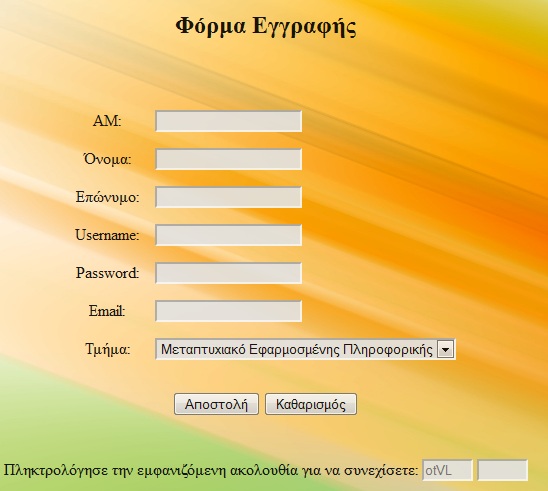}
	\caption{Registration Form for guest users}
\end{figure}

The guest users can only see the name of lecture files, apply for registration to the platform and finally communicate with the administrator. The system has been designed in a way such that not everyone can access the platform. Only the Administrators have the ability  to allow the users to have access. When a user wants to have access to the system he fills  his personal details in a form.  The application makes use of  a captcha textbox such that to deny access to bots that make automatic registrations . Also there is  a check for duplicates  not only for the personal details but also for the  e-mail. If all these  informations  are unique  the administrator  grants the user the ability to use the platform. Also, at this point if a user wants to recover his password, he can complete his username and the new password will be sent to his e-mail address. 

\begin{figure}[!h]
	\centering
	\includegraphics[scale=0.3]{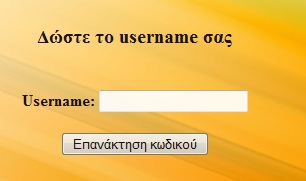}
	\caption{Password recovery}
\end{figure}

In addition  there is a second group of users, the registered users. In this group they can view and edit some of their account information, can read the lecture files, do the lecture tests in order to test their knowledge and communicate with other users or the trainer through a chat module .Finally they can take exams that are defined by the admin.

\begin{figure}[!h]
	\centering
	\includegraphics[scale=0.3]{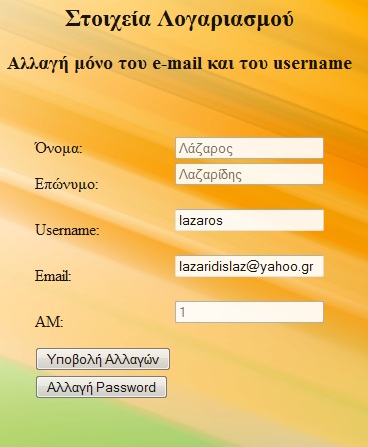}
	\caption{Account information of a registered user}
\end{figure} 

As shown in figure 3, a registered user can edit his username, e-mail and password. The registration index, the name and the last name  cannot be edited by the user, but only from the administrator. One of the most important module of the application is the tests and exams procedure. All the questions appear in random order from a pool a large number of questions that are exist in the database. So if someone takes the first lecture test more than once, he might not even have one common question. Furthermore, if two users take the same test  simultaneously, either the questions don’t appear in the same order or they aren’t the same. Moreover, in multiple choice questions the answers appear in random order, with the first one always selected. The check of the answers is done automatically and the results are stored in the database. 
 
\begin{figure}[!h]
	\centering
	\includegraphics[scale=0.3]{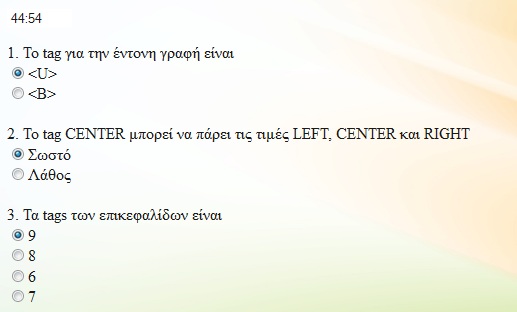}
	\caption{Exams of a registered user with multiple choice questions}
\end{figure}

\begin{figure}[!h]
	\centering
	\includegraphics[scale=0.3]{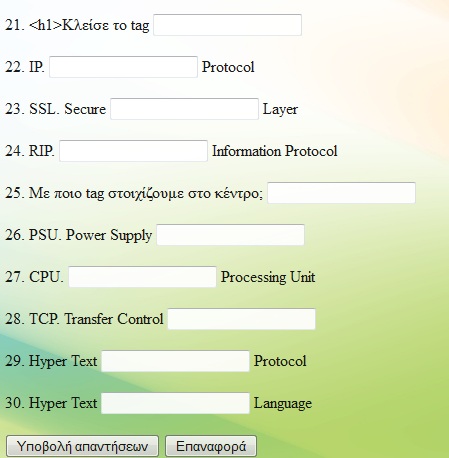}
	\caption{Exams of a registered user with gap\-filling questions}
\end{figure}

The registered user can communicate with the administrator via a contact form. The day/time, name and e-mail of the user  are completed automatically by the system and all he has to do is to  compose the message.

\begin{figure}[!h]
	\centering
	\includegraphics[scale=0.3]{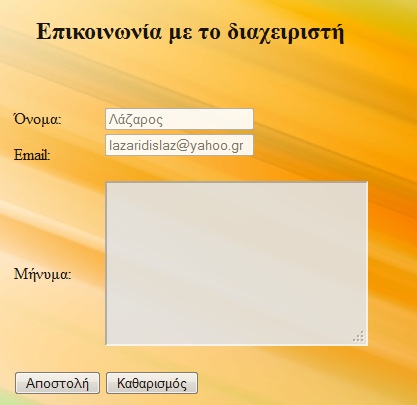}
	\caption{Communication of a registered user with the administrator}
\end{figure}

The third and last group of users are the administrators. They can insert, modify and delete users, lectures, as well as all the types of questions such as gap filling and/or multiple choice questions. They can also communicate with other users with chat or email, define the date of the tests and exams, as well as their duration. After the date/ time duration is defined , the users are informed automatically via email.

\begin{figure}[!h]
	\centering
	\includegraphics[scale=0.3]{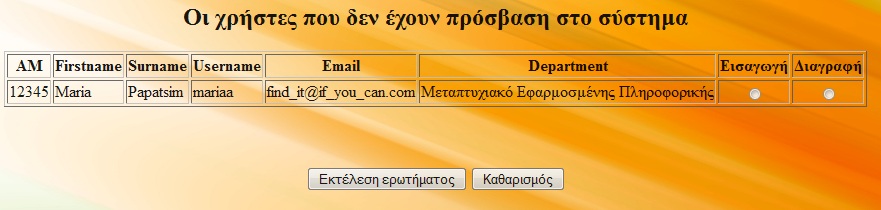}
	\caption{Insert / delete users who asked for permission}
\end{figure}

When the users decide to be members of the platform and apply for , they don’t immediately obtain access. The administrator must authenticate that they are really community “members” and then either grant access to them or delete them from the system minimizing the possibility of "bots" scripts .

\begin{figure}[!h]
	\centering
	\includegraphics[scale=0.3]{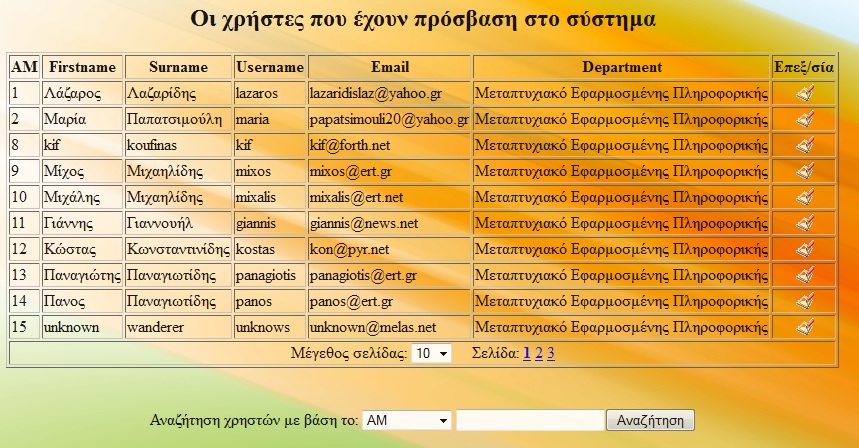}
	\caption{Edit details of registered users }
\end{figure}

Moreover, the administrator is able to view all the registered users and their personal details, but he can edit only some of them such as their Register Number, last name and name. In addition, he has the potential to search for registered users based on their Register Number, name, last name, username or e-mail by inserting a  string of characters. The results are displayed based on the category chosen.  In the “Delete Users” section, the users are displayed in order based on their Register Number and the administrator is able to delete them by choosing the corresponding checkbox.

\begin{figure}[!h]
	\centering
	\includegraphics[scale=0.3]{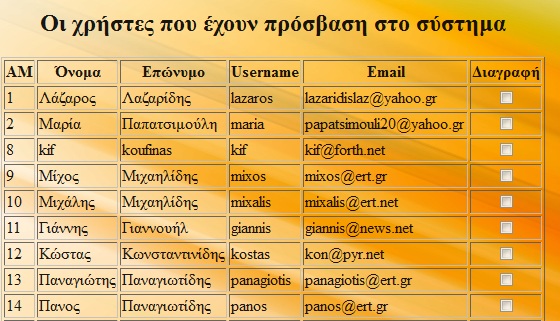}
	\caption{Deletion of Users}
\end{figure}

The administrator has also the ability to insert, delete and edit the lectures.

\begin{figure}[!h]
	\centering
	\includegraphics[scale=0.3]{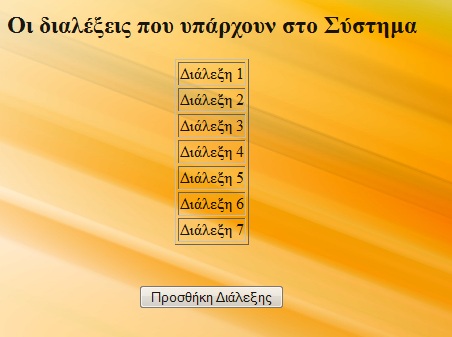}
	\caption{Insert Lectures}
\end{figure}

Due to the fact that the lectures that can be taken in each course vary, the administrator has the ability to add lectures adjusted to the needs of each course.
The opposite procedure is the deletion of the lectures. During that, not only the lecture files but also the questions related to the specific lecture are deleted.

\begin{figure}[!h]
	\centering
	\includegraphics[scale=0.3]{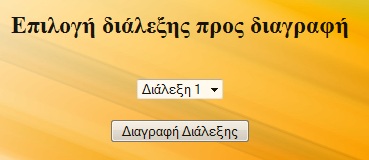}
	\caption{Lecture selection for deletion}
\end{figure}

Furthermore, the administrator is able to display, upload and delete files of each lecture (figure 12).  

\begin{figure}[!h]
	\centering
	\includegraphics[scale=0.3]{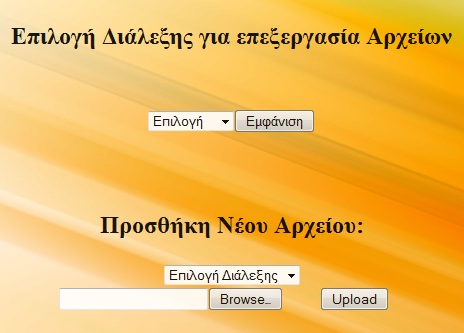}
	\caption{File processing}
\end{figure}

He can also select the lecture whose files need to be processed. In figure 13 the name, size and file type of the lecture files are shown.

\begin{figure}[!h]
	\centering
	\includegraphics[scale=0.3]{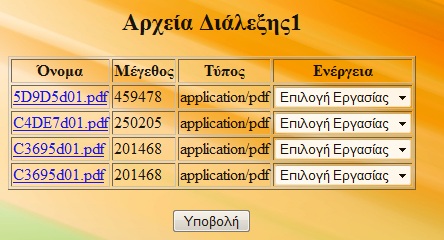}
	\caption{Deletion of lecture1' files}
\end{figure}

In the column “Action” the administrator can select which files he wishes to delete from a specific lecture. By pressing the “Submit” button, the files that have been marked as “Delete” will be erased from the database. Another important module of the application that the administrator has access is the questions category. The questions in the platform  are divided in multiple choice questions and gap filling questions. During the insertion of the questions according to the type of question  the form of the fields changes. For instance, if the type of question that has been chosen by the administrator is a gap filling question, the result will be as shown in figure 14.

\begin{figure}[!h]
	\centering
	\includegraphics[scale=0.3]{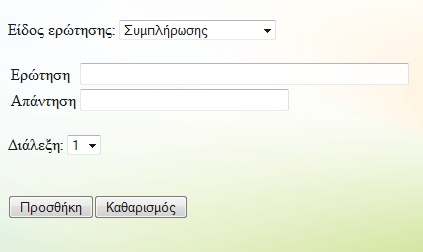}
	\caption{Insert filling question}
\end{figure}

The administrator is asked to fill in both the question and the right answer, but also to choose in which lecture the specific question belongs. In case that the administrator chooses to insert a multiple choice question, the menu is adjusted as shown in figure 15.

\begin{figure}[!h]
	\centering
	\includegraphics[scale=0.3]{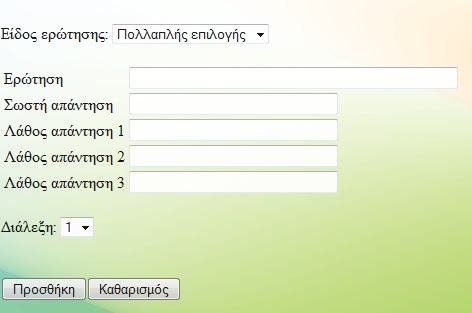}
	\caption{Insertion of a multiple choice question}
\end{figure}

The administrator is asked to choose in which lecture the question belongs and fill in the fields of the question, the right answer and at least of one wrong answer. The administrator can also delete one or more questions using a menu. 

\begin{figure}[!h]
	\centering
	\includegraphics[scale=0.3]{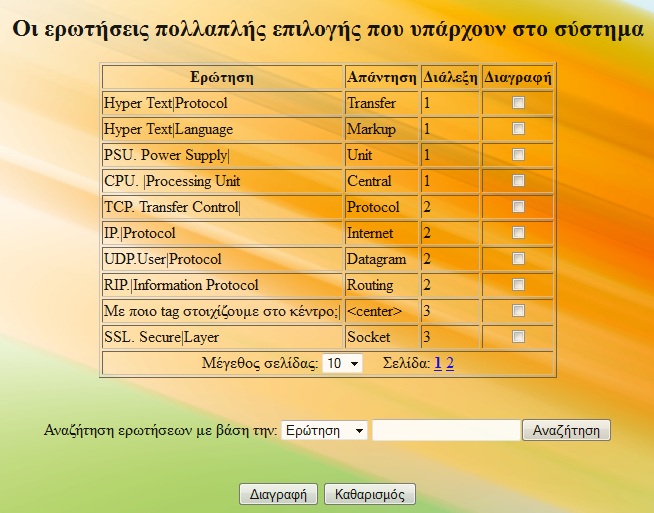}
	\caption{Gap filling question deletion}
\end{figure}

During the view of gap-filling questions for deletion, the questions are ordered according to the lecture. The fields of the question, the answer and the lecture in which they belong are displayed, as well as the checkbox that is chosen for question deletion. The administrator can change the page size, he can also search for some question by entering  a string in the textbox. The search is executed either there is a full search string or a part of the word we are looking for  . The same procedure also applies to multiple choice and gap-filling questions as well .

\begin{figure}[!h]
	\centering
	\includegraphics[scale=0.3]{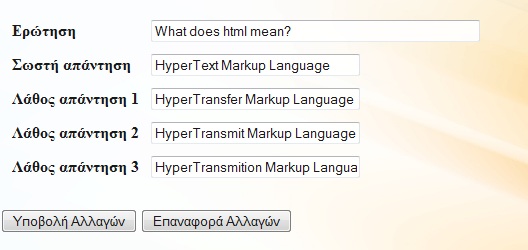}
	\caption{Multiple choice question processing}
\end{figure}
 
During processing the administrator can make changes to the questions and to the answers. Administrator also has the ability to send massive e-mails to all registered users.

\begin{figure}[!h]
	\centering
	\includegraphics[scale=0.3]{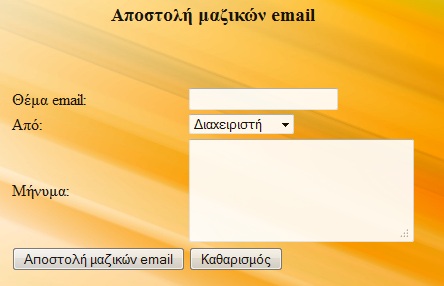}
	\caption{Massive e-mail sending}
\end{figure}

Finally, the administrator sets the dates and the duration of the tests / exams. The users are informed via e-mail for any changes automatically .

\begin{figure}[!h]
	\centering
	\includegraphics[scale=0.3]{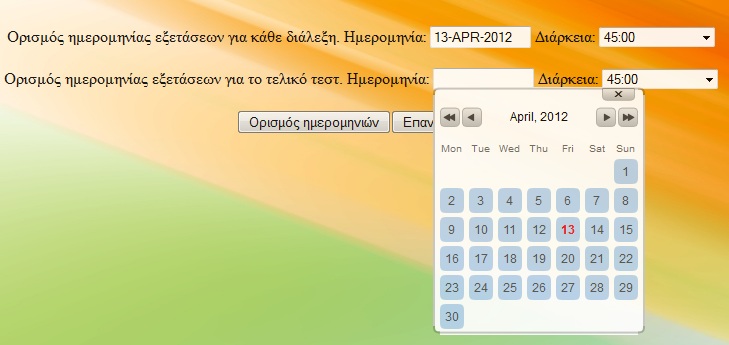}
	\caption{Setting the date and the duration of the tests and exams}
\end{figure}

\section{MySQL and Database Architecture}

MySQL is the most well-known Database Management System \cite{mysql}. It is an open source tool and its major advantage is that it is constantly improved in regular time intervals. As a result MySQL has evolved into a fast and extremely powerful database. It performs all functions like storing, sorting, searching and recall of data in an efficient way. Moreover, it uses SQL, the standard worldwide query language. Finally, it is distributed free. The Database which is used in the application has been designed to use the storage engine InnoDB, so that the restrictions of the foreign keys can be created. It consists of twelve tables, eight of which are the most important as they are connected to each other with relations. Its complete diagram is as follows: 

\begin{figure}[!h]
	\centering
	\includegraphics[scale=0.3]{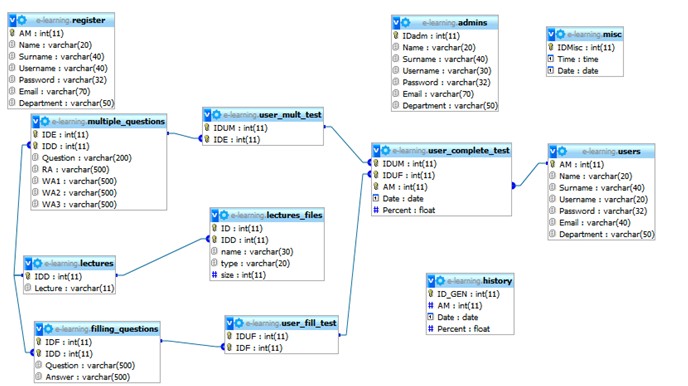}
	\caption{The platform’s database scheme}
\end{figure}

The table register consists of: field AM (Type: integer) which is the primary key. The fields Name, Surname, Username, Password, Email and Department are varchar type. In this table the records which represent the users’ applications in order to access the system, are stored.  The table admins consists of: the field IDadm (Type: integer) which is the primary key. The fields Name, Surname, Username, Password, Email and Department are varchar type. In this table all the administrators of the system, are stored. The table misc consists of fields IDMisc (Type: integer) which is the primary key, Time (Type: time, format: hh:mm:ss) and Date (Type: date, format: mm\-yyyy\-dd). In this table is stored the date and the time that the final and lecture tests will take place. 
The table history consists of fields ID\_GEN (Type: integer) which is the primary key, AM (Type: integer), Date (Type: date, format: mm\-yyyy\-dd) and Percent (Type: float). In this table are stored all the final tests that the user gave, with every date and percentage he achieved each time.  In the table lectures are stored only the lectures of the lessons. It consists of IDD (Type: integer) which is the primary key and Lecture (Type: varchar). In table lecture\_files are stored the file paths that are related with lectures. It consists of ID (type: integer), IDD (Type: integer), name (Type: varchar), type (Type: varchar) and size (Type: integer). The primary key is ID whereas IDD is the foreign key which defines in which lecture the file of the lesson is referred. In the filed name is stored the full path of the file. In the field type is stored the type of the file whereas in the field size its size.    In tables multiple\_questions and filling\_questions are stored multiple choice questions and filling questions respectively. The fields of multiple\_questions table are IDE (Type: integer), IDD (Type: integer) whereas Question, RA, WA1, WA2, WA3 are varchar type. IDE is the primary key whereas IDD is a foreign key which defines in which lecture each multiple choice question, belongs. The field Question contains the question, RA contains the right answer and in the rest three fields are the wrong answers.  The fields of the table filling\_questions are IDF (Type: integer), IDD (type: integer), Question (Type: varchar) and Answer (Type: varchar). IDF is the primary key whereas IDD is a foreign key which defines in which lecture each filling question, belongs. In the field Questions the question is stored, whereas in the field answer the right answer. Each of user\_mult\_test and user\_fill\_test tables consist of two fields, where both of these fields are keys. Fields IDUM (Type: integer) and IDUF (Type: integer) are the primary keys respectively. These two tables are intermediate stops for the creation of the final table user\_complete\_test that is shown below. Their presence is necessary and the reason is, that for the various tests which the system provides the number of questions for each type is different. In the final test of the system for instance are chosen twenty multiple choice questions and ten filling questions. Supposing we want to create a test with ten multiple choice and five gap\-filling questions. 

\begin{figure}[!h]
	\centering
	\includegraphics[scale=0.3]{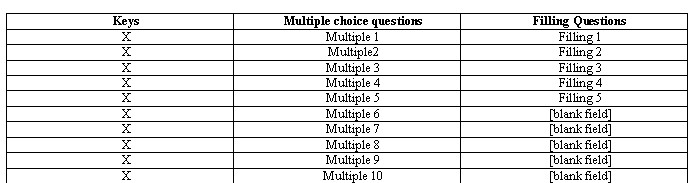}
	\caption{Example of 10 multiple choice and 5 gap filling questions in one table}
\end{figure}

The questions are grouped in the user\_multr\_test and user\_field\_test tables and each group is put as a foreign key in the user\_complete\_test\_table .   In the table users are stored the registered users of the system. It consists of the fields AM (type: integer) which is the primary key and Name, Surname, Username, Password, Email, Department which are varchar. The fields are equally important with fields of the register table.    In the final user\_complete\_test table are stored all the tests that the user gave, either they refer to lecture tests or final tests. Its fields are IDUM (Type: integer), IDUF (type: integer), AM (Type: integer), Date (Type: date, format: mm\-yyyy\-dd) and Percent (Type: float). The table doesn’t have a primary key of its own, but three foreign keys (IDUM, IDUF, AM) are its primary. In the field AM is stored the record number of the student to whom the test corresponds. In the fields IDUM and IDUF are stored the multiple choice question and filling groups that have been chosen for the specific AM test. In the field Date is stored the date that the test is held and in the field Percent the percentage that the student achieved.

\section{Conclusion}
 In the present paper the  S.A.T.E.P. : Synchronous-Asynchronous Tele-education Platform  for educational purposes is developed and can be easily parametrized  to meet the needs for any educational Inst. The administrator can  upload  the files that the user intends to read, but also inserts the questions of each lecture. The trainee can read/download the files of each lesson and also test his knowledge on what he has studied through a series of questions/tests. The automatic mix of the exam questions and their automatic correction as well, makes the current assignment different from the rest of the Content Management Systems (CMS). Unlike CMS, the current platform is ready to be used   and doesn’t need the addition of extra plugin.

\end{document}